# *In-situ* study of oxygen exposure effect on spin-orbit torque in Pt/Co bilayers in ultrahigh vacuum

Hang Xie,[1] Jiaren Yuan,[2] Ziyan Luo,[1] Yumeng Yang,[1] and Yihong Wu[*,1]

[1]Department of Electrical and Computer Engineering, National University of Singapore, Singapore 117583, Singapore

[2]College of Science, Jiangsu University, Zhenjiang, 212013, China

Oxygen incorporation has been reported to increase the current-induced spin-orbit torque in ferromagnetic heterostructures, but the underlying mechanism is still under active debate. Here, we report on an *in-situ* study of the oxygen exposure effect on spin-orbit torque in Pt/Co bilayers via controlled oxygen exposure, Co and Mg deposition, and electrical measurements in ultrahigh vacuum. We show that the oxygen exposure on Pt/Co indeed leads to an increase of spin-orbit torque, but the enhancement is not as large as those reported previously. Similar enhancement of spin-orbit torque is also observed after the deposition of an MgO capping layer. The results of *ab initio* calculations on the Rashba splitting of Pt/Co and Pt/Co/O suggest that the enhancement is due to enhanced Rashba-Edelstein effect by surface-adsorbed oxygen. Our findings shed some light on the varying roles of oxygen in modifying the spin torque efficiency reported previously.

* E-mail: elewuyh@nus.edu.sg

## Introduction

Spin-orbit torque (SOT) induced by an in-plane charge current in ferromagnet (FM) / heavy metal (HM) multilayers has drawn considerable interest due to its great potential in magnetization switching without specific requirement on the magnetization configuration in the current path[1,2]. It is commonly believed that the SOT in FM/HM bilayers is induced by two major mechanisms: one of them is the spin Hall effect (SHE) in the bulk of the HM layer[3-7], and the other is the Rashba-Edelstein effect (REE) at the FM/HM interface due to structure inversion asymmetry[8-11]. Depending on the type of FM and HM materials and their thicknesses or even the capping or underlayer, one of the mechanisms may become dominant over the other, but it is now commonly accepted that both mechanisms are present in FM/HM bilayers or oxide/FM/HM trilayer structures. Both the SHE and REE can induce damping-like (DL) and field-like (FL) torques exerting on the magnetization of the FM layer. The strength of the DL and FL torques, or the corresponding effective fields, can be quantified through the magnetization dynamics induced by the current. From application point of view, a large current to torque conversion efficiency is highly desirable as this would lead to low energy consumption. In this context, various attempts have been made to increase the SOT generation efficiency ever since it was first observed in Pt/Co/AlO$_x$ trilayers[9]. The approaches taken so far include but are not limited to the use of different FM and HM materials[12,13], optimization of FM and HM deposition processes as well as their thicknesses[14,15], the use of HM with opposite spin Hall angle at both sides of the FM[16], or composite HM made of different materials[17], *etc*. The effects of these structural or material changes on SOT can be accounted for to a reasonable degree of success by the simple drift-diffusion formalism or Rashba effect[18].

In addition to the aforementioned approach of using different materials, attempts have also been made to manipulate the SOT by incorporating oxygen either inside the constituent materials or at the interfaces. Several groups have found that, by incorporating oxygen in (Co,

CoFeB)[19-22] and (Pt, W)[23-25] layer, or by controlling the degree of oxidation in light metal layer (Mg, Al, Cu)[26-28], it is possible to induce sizable changes in the SOT. When oxygen is in the vicinity of the interfaces, *e.g.*, in Co(O)/GdO$_x$[19], Co(O)/HfO$_2$[21], CoFeB(O)/MgO[26], FM/HM (Co(O)/Pt[22] and W(O)/CoFeB[25]), the increase of SOT is attributed to the enhancement of REE, which in turn is believed to be caused by the modulation of work function or electric potential at the interfaces due to oxygen incorporation. Similar mechanism was also proposed to account for the SOT observed in NiFe/Al$_2$O$_3$ and NiFe/CuO$_x$[27,28], in which the HM is replaced by oxide of metals with small spin-orbit coupling. On the other hand, in the case of Pt(O)/NiFe/SiO$_2$ with heavily oxidized Pt, the observed dominant DL SOT is attributed to an intrinsic SOT arising from the Berry curvature at the interface[23]. In addition to conventional REE, the combined effect of enhancement in orbital moment and spin-orbit coupling was also proposed to account for the sign reversal of DL SOT in Pt/CoFeB(O)/MgO below temperature of 200 K[20], and significant enhancement in DL torque in Pt/Co/SiO$_2$[29]. The oxygen may also modify the charge and spin transport property at the interface, *e.g.*, it was reported that spin-mixing conductance of Co/CoO$_x$/Pt is enhanced due to antiferromagnetic nature of CoO$_x$[21,22], and the spin Hall angle in Pt is increased by weakening of proximity effect[22]. Apart from interface-related effects, it is also demonstrated that oxygen incorporation could affect the spin-dependent scattering in the bulk of naturally oxidized Cu and lift the spin Hall angle of Cu to a value comparable to that of HM[28].

Based on the above, it is apparent that currently there is no consensus regarding both the effect and underlying mechanism of oxygen on SOT. In all the experiments reported so far, oxygen is incorporated in the form of oxide, and therefore, it is difficult to ascertain if the effects observed are due to the oxide layer in direct contact with the ferromagnet or atomic/molecular oxygen. This is important because in our previous studies we have unveiled that the effects of oxygen and oxide are different when it comes to the influence on perpendicular magnetic anisotropy (PMA)[30]; oxide capping layer ( *e.g.*, MgO) enhances PMA of Co whereas oxygen adsorbed on the surface of Co brings about the different effect, *i.e.*, decreasing the magnetic moment. Therefore, it would be of interest to know if similar effects are present for SOT when the sample is either exposed to oxygen or capped by an oxide. In addition, the use of oxide makes it difficult to investigate the effect of oxygen on the same sample by varying the oxygen dose systematically because once the oxide is formed the only way to vary the oxygen dose and profile is by thermal annealing. Although thermal annealing can change the oxygen distribution inside the sample within a certain range, it is difficult to control the oxygen dose and distribution in a quantitative manner. In this context, in this work, we conducted a well-controlled oxygen exposure study of spin-orbit torque in Pt/Co heterostructures in an *in-situ* setup. The *in-situ* deposition/characterization capability in an ultrahigh vacuum (UHV) environment allows precise control of both the dose and location of oxygen at sub-monolayer accuracy, and importantly, quantification of the SOT measurements can be carried out after each oxygen exposure without breaking the vacuum; this enables us to examine the oxygen effect in a well-controlled fashion, which was not possible in previous work. In contrast to large enhancement reported so far, we found that the oxygen exposure of Co leads to a relatively small enhancement of SOT efficiency in Pt/Co bilayers at low oxygen dose, and the SOT starts to saturate with further increasing the oxygen dose. On the other hand, the deposition of a thin MgO layer on top of the Co layers also enhances the SOT. By performing *ab initio* calculations, we found that the Rashba splitting is weak in pristine Pt/Co bilayers, while with oxygen adsorption the Rashba splitting is evidently enhanced. This enhancement may be resulted from the large local electric field at Co/O surface, which leads to enhancement of SOT observed in the oxygen-exposed Pt/Co. These results demonstrate the importance of oxygen in modifying the surface and interface states, thereby enhancing the SOT.

## Methods

As we mentioned above, as ultrathin Co can be oxidized easily, it is impossible to characterize oxygen effect on SOT of Pt/Co bilayers under ambient condition. This motivated us to carry out all the experiments *in-situ* in an UHV environment. As shown in Figure 1a, the *in-situ* characterization system consists of a load-lock chamber, a nano-probe chamber with a scanning electron microscope, four nano-probes and an electromagnet, and a deposition chamber. All the chambers are connected by a transfer chamber with a base pressure of $9 \times 10^{-10}$ mbar. The base pressures of the nano-probe and deposition chambers are $2 \times 10^{-10}$ and $1 \times 10^{-9}$ mbar, respectively. More details about the UHV system can be found elsewhere[31,32]. To facilitate *in-situ* electrical measurements, Hall bars with a dimension of 15 μm × 30 μm were first formed on thermally oxidized Si substrates before the deposition of thin films. Since the *in-situ* deposition chamber is only equipped with Knudsen cells, the Ta seed layer (3 nm in thickness) and Pt layer (2 nm in thickness) were deposited *ex-situ* using a separate sputter. After the Ta/Pt deposition, the samples were immediately fixed on a magnetic holder which is able to provide either an out-of-plane field or in-plane field (maximum 1500 Oe) and loaded into the UHV chamber. The samples were transferred back and forth between the nano-probe and deposition chamber through the transfer chamber without breaking the vacuum. The nano-probe chamber is equipped with a precision oxygen source which allows to expose the sample surface to oxygen at sub-monolayer (nominal coverage of oxygen at unity sticking coefficient) accuracy by simply controlling the oxygen partial pressure from the chamber base pressure of $2 \times 10^{-10}$ mbar to $1.5 \times 10^{-8} - 4 \times 10^{-7}$ mbar and the exposure duration between 10 m and 3 h. The deposition of Co and Mg was done *in-situ* using K-cells in the preparation chamber, with a rate of 0.033 Å s$^{-1}$ and 0.265 Å s$^{-1}$, respectively. Prior to deposition of any films or oxygen exposure, the sample was preheated in the deposition chamber for an hour at 110°C to remove moisture, if any, on the surface of the Hall bar which was already deposited with Ta/Pt layers.

The SOT measurements were carried out in the nanoprobe chamber. Standard harmonic Hall voltage measurement is employed to characterize the SOT[33], and the Hall voltage is measured using two voltage electrodes.

Density function theory (DFT) calculations for Pt/Co bilayers were performed by using the Vienna *ab initio* simulation package (VASP) with the pseudopotentials and wave functions described by the projected augmented-wave method[34]. The exchange correlation energy is calculated within the generalized gradient approximation in the form of Perdew-Burke-Ernzerhof[35]. The cut-off energy for the plane-wave basis set used to expand the electron wave functions is set to be 450 eV, with the Γ-centred Brillouin zone sampled by a 25×25×1 *k*-point mesh and the convergence criterion of the total energy set to be $10^{-6}$ eV. The calculation is performed in three steps. First, structural relaxations were carried out until none of the forces exceeded 0.01 eV Å$^{-1}$ to determine the most stable structure geometries. Next, the Kohn-Sham equations were solved with a collinear calculation method without spin-orbit coupling (SOC) to find out the charge distribution of the ground state. Finally, SOC was taken into account and the self-consistent total energy of the system was obtained with different spin orientations.

## Results and Discussion

The inset of Figure 1a shows the schematic of a Hall bar used for SOT characterization and the sample structure. The layer sequence of the sample is as follows: Si substrate/SiO$_2$/Ta (3)/Pt (2)/Co (*t*), where the number inside the parentheses indicates the layer thickness in nm (from now onwards, the Ta seed layer is omitted for simplicity). As we reported previously[30], Pt (2)/Co (*t*) exhibits in-plane magnetic anisotropy (IMP) when *t* = 0.2 – 0.5 nm and clear perpendicular magnetic anisotropy (PMA) when *t* is in the range of 0.6 – 1.2 nm beyond which it goes back to IMP. Since this work is concerned about Co films with PMA, we set *t* at 0.8 nm.

After the deposition of 0.8 nm Co on the pre-heated Pt layer, the sample was transferred to the nano-probe chamber for electrical measurements. Figure 1b shows the AHE curve of the pristine Pt (2)/Co (0.8) sample, where the *y*-axis is the Hall resistance which is proportional to the *z* component of magnetization. As can be seen from the AHE curve, a well-defined PMA is formed in Co at a thickness of 0.8 nm. Subsequently, we exposed the sample to oxygen using the *in-situ* oxygen source by gradually increasing the dose from 5.5 L to 300 L (here, L is the Langmuir unit with 1 L corresponding an exposure of $1.33 \times 10^{-6}$ mbar for 1 s). If we assume a unity sticking coefficient, a full coverage of the energy favourable *hcp* hollow sites on Co surface requires an oxygen exposure dose of about 5.2 L, and therefore 5.5 L is presumably sufficient to cover the entire Co surface[30]. During the process, the oxygen exposure is paused at selected doses to allow electrical measurements.

From the measured AHE loops at different doses, we can estimate the saturation magnetization normalized to pristine values as $M_s/M_s^0 = \Delta R_{AHE}/\Delta R_{AHE}^0$, where $M_s^0$ ($M_s$) and $\Delta R_{AHE}^0$ ($\Delta R_{AHE}$) are the saturation magnetization and anomalous Hall resistance of Pt (2)/Co (0.8) without (with) oxygen exposure, respectively. This is possible because the longitudinal resistance remains almost the same after oxygen exposure and thus the change in AHE caused by the oxygen effect on electrical property is presumably small. As shown in Figure 1c, $M_s$ of the bilayer drops monotonically with the oxygen dose. The dropping is fast below 30 L, after which it gradually slows down as oxygen dose increases. This is in good agreement with our previous findings[30], and is attributed to partial quench of the magnetic moment of the surface layer by the charge transfer from Co to the adsorbed oxygen. The decrease in dropping rate at larger dose is due to the self-limiting process of oxygen adsorption on Co at room temperature. In addition to $M_s$, we have also investigated the oxygen effect on PMA. From the AHE loops, the bilayer maintains PMA within the maximum oxygen dose range, though the coercivity keeps decreasing as oxygen dose increases. As the field strength of the electromagnet in the UHV system is insufficient to saturate the magnetization in-plane, we have tried to extract the effective anisotropy field $H_k^{eff}$ via fitting the $V_{AHE} - H_x$ loop using the relationship $V_{AHE} = V_m \left[1 - \frac{1}{2}\left(\frac{H_x}{H_k^{eff}}\right)^2\right]$ at small field (where $V_m$ is the amplitude of anomalous Hall voltage and $\theta$ is the polar angle of magnetization). The results showed that $H_k^{eff}$ maintains around 1.3 T within the oxygen exposure range (*i.e.*, < 300 L). These indicate that the oxygen exposure within this dose range merely reduces the magnetic moment without degrading the PMA of Pt (2)/Co (0.8).

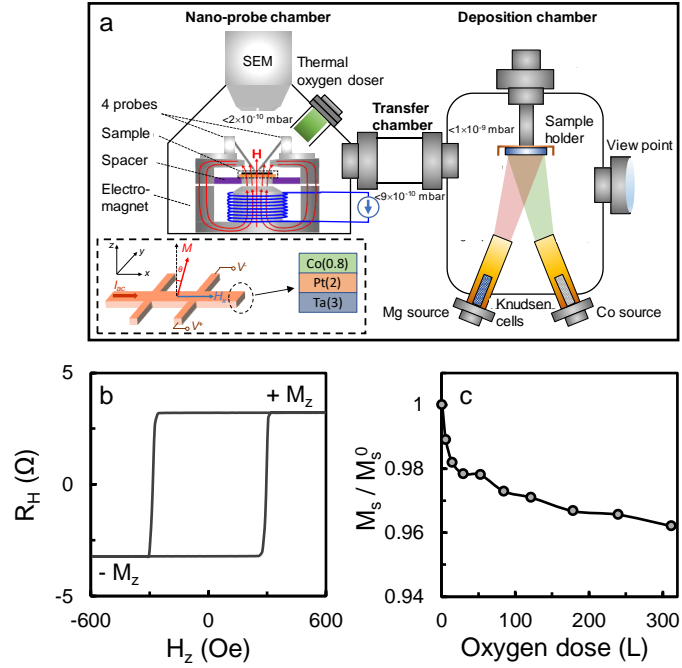

Figure 1. (a) Schematic of the UHV system for *in-situ* characterization and deposition, which consists of a nano-probe chamber, a transfer chamber and a deposition chamber. The inset shows the schematic of Hall bar used for SOT characterization and the sample structure. (b) AHE loop for Pt (2)/Co (0.8) measured with a perpendicular magnetic field $H_z$, where $+M_z$ and $-M_z$ indicate the up and down magnetization. (c) Normalized saturation magnetization $M_s/M_s^0$ versus oxygen dose.

The above results demonstrate that the oxygen effects on PMA and $M_s$ reported previously are reproducible. Next, we discuss the oxygen effect on SOT measured using the harmonic Hall voltage method. To this end, an AC current is applied to the Hall bar, and the first and second harmonic Hall voltages are measured simultaneously as a function of the in-plane bias field: $H_x$ for measuring DL SOT and $H_y$ for measuring FL SOT. Since the direction of magnetic field provided by the in-plane magnetic holder cannot be changed, we fixed two samples with the same structure on the magnetic holder, whose longitudinal directions are parallel and perpendicular to the magnetic field for measuring DL and FL SOT, respectively. The similarity in transport properties of the two samples was confirmed before SOT characterization. The SOT effective field $H^{DL(FL)}$ is obtained from the respective curvature and slope of $V^\omega$ and $V^{2\omega}$ versus the in-plane field as[36]

$$H^{DL(FL)} = -2\left(\frac{\partial V^{2\omega}}{\partial H_{x(y)}} \bigg/ \frac{\partial^2 V^\omega}{\partial H_{x(y)}^2}\right). \tag{1}$$

Figure 2a and 2b show the first harmonic Hall voltages versus in-plane field $H_x$ and $H_y$ of Pt (2)/Co (0.8) at an AC current amplitude of 2.4 mA (corresponding to a current density of 5.33 × 10$^6$ A/cm$^2$ calculated by assuming 67% of the current flows in the 2 nm Pt layer based on the calibrated resistivity). Due to the small nucleation field of the sample, we fit the first order voltage from -1000 Oe to 200 Oe for down magnetization state and from -200 Oe to 1000 Oe for up magnetization state, where there is no nucleation occurring. As shown, the first order voltage shows a parabolic shape with the in-plane field. Figure 2c and 2d show that the second order voltage is liner with in-plane field. Additionally, it can be seen that the direction of $H^{DL}$ reverses as the magnetization changes direction while the direction of $H^{FL}$ is independent of the magnetization direction. The measurement is conducted at different AC current amplitudes from 1.6 mA to 3.2 mA. $H^{DL}$ and $H^{FL}$ are extracted from the measurement data using Equation (1).

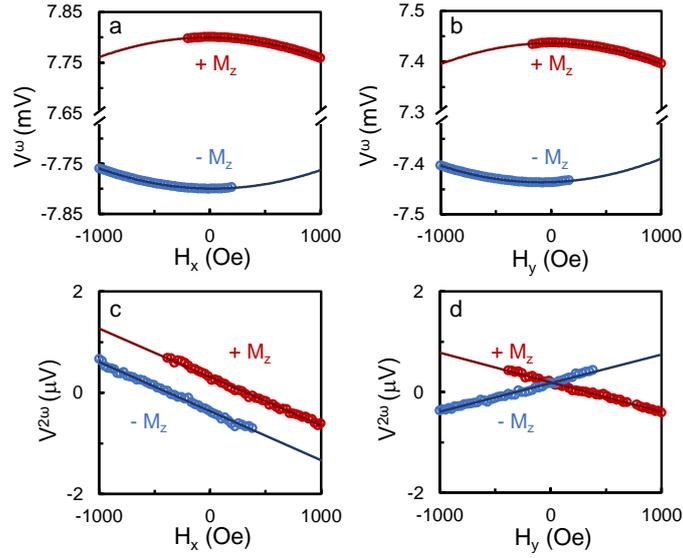

Figure 2. First harmonic Hall voltages of Pt (2)/Co (0.8) with applied in-plane field of (a) $H_x$ and (b) $H_y$, and second harmonic Hall voltages with applied in-plane field of (c) $H_x$ and (d) $H_y$ measured at an AC current with an amplitude of 2.4 mA. + $M_z$ and - $M_z$ indicate the up and down magnetization.

The extracted $H^{DL}$ and $H^{FL}$ are shown in Figure 3a and 3b respectively as a function of current density for three different oxygen doses. As can be seen, $H^{DL}$ and $H^{FL}$ increase linearly with $J$ and the slope is different for different oxygen doses. We have repeated the same experiments at other oxygen doses and from the slopes of $H^{DL}$ and $H^{FL}$ versus $J$ curve, we obtained $H^{DL}/J$ and $H^{FL}/J$ at different oxygen doses and the results are shown in Figure 3c and 3d, respectively. As can be seen, both $H^{DL}/J$ and $H^{FL}/J$ initially increase relatively quickly with the oxygen dose and then the increase slows down and gradually saturates at larger oxygen dose, a trend which is similar to saturation magnetization shown in Figures 1c, though the polarity is opposite.

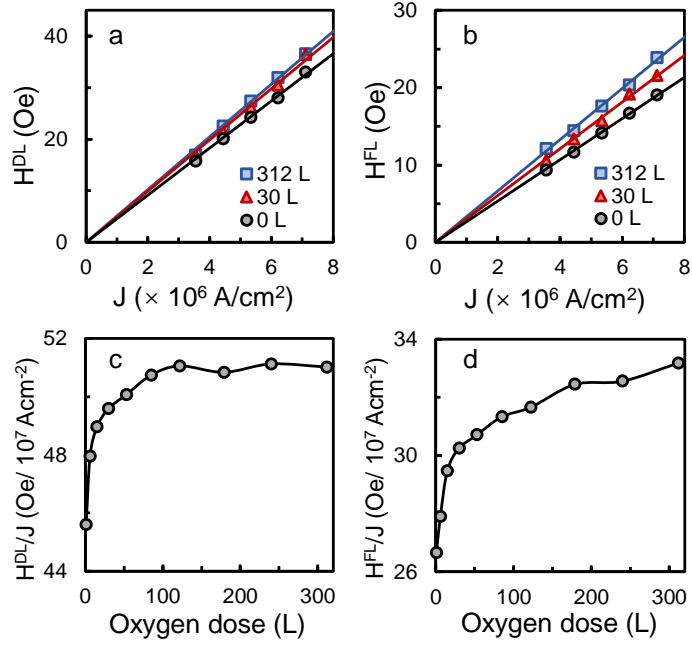

Figure 3. (a) Damping-like SOT effective field $H^{DL}$ and (b) field-like SOT effective field $H^{FL}$ as a function of current density $J$ at oxygen dose of 0 L (circle), 30 L (triangle) and 312 L (square); the lines are linear fittings of $H^{DL}$ versus $J$. (c) $H^{DL}/J$ and (d) $H^{FL}/J$ as a function of oxygen dose.

At the early stage of SOT studies, the DL and FL torques are thought to be originated from different effects, *i.e.,* the former is mainly from bulk SHE and the latter from interfacial REE. But, it is now commonly accepted that both the SHE and REE can contribute to the DL and FL torques. Therefore, the DL and FL SOT efficiencies $\xi^{DL(FL)}$ can be estimated from[37]

$$\xi^{DL(FL)} = \frac{2e}{\hbar} M_s t_{Co} \frac{H^{DL(FL)}}{J}, \qquad (2)$$

where $M_s$ is the magnetization, $\hbar$ the Plank constant, $e$ the electron charge, and $t_{Co}$ the Co thickness (fixed at 0.8 nm). Using an $M_s^0$ value of 1400 emu/cm³ and the values of $M_s/M_s^0$ estimated from $\Delta R_{AHE}/\Delta R_{AHE}^0$, as well as values of $H^{DL}/J$ and $H^{FL}/J$ shown in Figures 3c and 3d, $\xi^{DL}$ and $\xi^{FL}$ for each oxygen dose are calculated using Equation (2) and the results are shown in Figure 4. The extracted values of $\xi^{DL}$ and $\xi^{FL}$ for pristine Pt (2)/Co (0.8) are 0.155 and 0.09, corresponding well with reported values of Pt/Co system in litertaure[19,38-40]. We noticed that the FL SOT observed in this study is much larger than some of the reported values in literature [3,41,42], which may be due to the relatively small Pt thickness used in the study. This is affirmed by the fact that, under same measurement conditions, the FL SOT of Pt (5)/Co (0.8) is around one order of magnitude smaller than that of Pt (2)/Co (0.8). This Pt thickness dependence of FL SOT might be related to relative contribution from the interface at different Pt thicknesses. Since there is no oxide capping layer on the pristine Pt/Co layers and the REE has been reported to be weak at Pt/Co interface[3,38], we believe that SHE is dominant in generating the SOTs in this system. Further studies are required in order to quantify the respective contributions of SHE and REE in generating the FL SOT in uncapped Pt/Co bilayer. Similar to the trends of $H^{DL}/J$ and $H^{FL}/J$ versus oxygen dose, both $\xi^{DL}$ and $\xi^{FL}$ experience initial fast increase below oxygen dose of 50 L, after which they increase slowly and gradually saturate, despite some slight difference between these two curves, *i.e.*, $\xi^{DL}$ increases by about 8 % when the oxygen dose increases from 0 L to 85 L beyond which it saturates at about 0.168 level with slight fluctuations, while $\xi^{FL}$ increases quickly at smaller dose and continues to increase at smaller rate till 310 L. Overall, the $\xi^{FL}$ increases by 18.5 %, which is larger than the 8 % increase of $\xi^{DL}$. Therefore, the enhancement of FL SOT efficiency is more evident than that of DL SOT, but the percentage of enhancement, even at maximum, is still much smaller than those reported in literature using the oxide capping layer[19,21,22]. We didn't observe any sign reversal at room temperature.

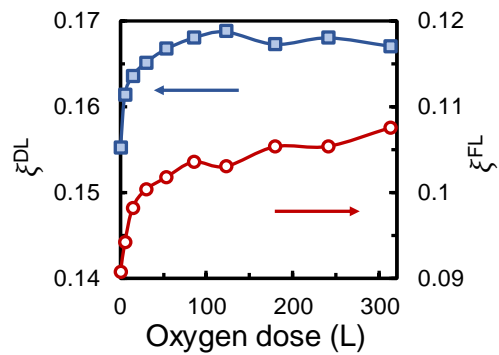

Figure 4. Damping-like SOT efficiency $\xi^{DL}$ (square, left axis) and field-like SOT efficiency $\xi^{FL}$ (circle, right axis) versus oxygen dose.

The increase of SOT observed in the above experiments appears to be consistent with the reported findings about the effect of oxygen incorporation in FM on SOT[19,21,22], which are interpreted as interface-related phenomena at HM/FM and FM/Oxide interfaces, *e.g.*, enhanced REE at the interface of FM and oxide[19,21,22], or efficient spin current transmission at Pt/Co interface due to antiferromagnetic nature of $CoO_x$[21,22]. However, in our case the oxygen is directly incorporated on bare Co surface without any capping layer followed by the *in-situ* SOT measurements conducted in UHV, and therefore, the enhancement of SOT is not related to Co/oxide interface. It cannot be explained by effect near Pt/Co interface either because it is very unlikely that the oxygen atoms can penetrate into the Co layer and reach the Pt/Co interface under the current experimental conditions, *i.e.*, the diffusion of oxygen to sub-surface Co layer is difficult at room temperature and relatively low oxygen dose due to the requirement of a large activation energy. A more likely scenario is that an oxygen-rich Co surface is formed upon oxygen exposure[43]. Considering the large electronegativity of oxygen, the most likely mechanism responsible for the enhancement of SOT induced by oxygen exposure is the modification of local electric field near the Co surface, which in turn causes Rashba splitting and thereby gives rise to additional SOT due to REE. In general, the Rashba interaction at the interface or surface is described as $H_R = \alpha_R(\boldsymbol{e_z} \times \boldsymbol{k}) \cdot \boldsymbol{\sigma}$[44], where $\alpha_R$ is the Rashba parameter, $\boldsymbol{e_z}$ is the direction of electric field (normal to the surface or interface), $\boldsymbol{k}$ is the electron momentum and $\boldsymbol{\sigma}$ is the Pauli matrix. Since $\alpha_R$ is proportional to the electric field strength $E$ in *z*-direction[44], it can be altered by changing $E$ either globally or locally. Both approaches have been reported to be able to tune the REE in a reasonably large range[45-48]. For instance, $\alpha_R$ is reported to increase from 0.02 eV Å to 0.14 eV Å with an applied electric field of 0.4 V/Å at polar $KTaO_3$ surface[46] and from 0.06 eV Å to 0.09 eV Å by an applied gate voltage of 1.5 V to -1 V in $In_{0.53}Ga_{0.47}As/In_{0.52}Al_{0.48}As$ heterostructures[47]. Compared to a global field applied to the sample, modifying electric field locally by surface modification has been reported to be more effective in enhancing the Rashba splitting, *e.g.*, uncovered Gd(0001) surface presents negligible Rashba splitting, whereas Gd(0001)/O surface shows evident Rashba splitting due to the enhancement of surface electric field[44].

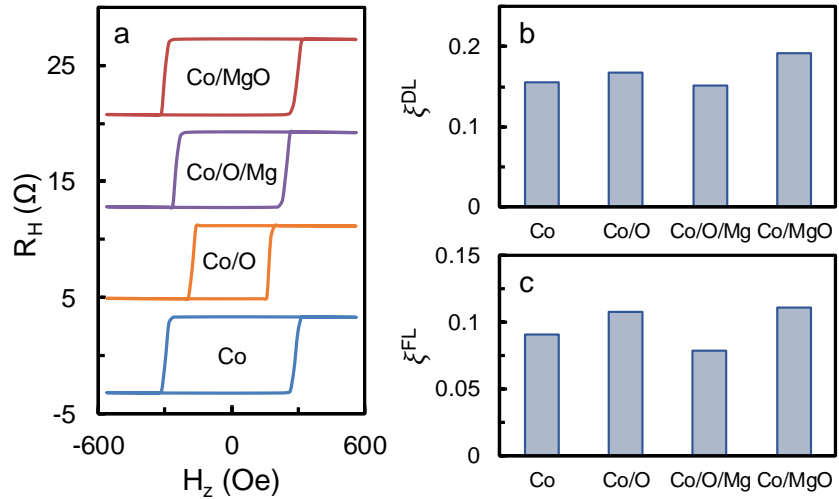

Figure 5. (a) AHE loops, (b) $\xi^{DL}$ and (c) $\xi^{FL}$ of Pt/Co, Pt/Co/O (312 L), Pt/Co/O (312 L)/Mg and Pt/Co/MgO, where Pt layer is omitted in the legend in the figures.

The present case is similar to the case of Gd/O[44]. After the oxygen exposure, the Co surface is covered by adsorbed oxygen atoms, though the exact coverage is unknown. Due to the large electronegativity of oxygen, some of the electrons will be transferred from Co to O[30,49], which results in a larger local asymmetry of charge distribution and thus increases the electric field near the Co surface. The increase of electric field at Co surface naturally leads to a larger Rashba splitting and an enhanced REE, which is directly reflected in the increased FL and DL SOT. It should be noted that at the very beginning of SOT studies, it is believed that the DL SOT is mainly due to the spin Hall effect. But, there is increasing evidence showing that REE can also give rise to DL SOT[9,10], which might result from the change of spin density caused by spin precession around the exchange field and spin flip in the Co layer[10]. The DL SOT generated from REE is expected to be much larger in system with ultrathin Co ferromagnetic layer than bulk Co due to much shorter spin diffusion length[10]. Therefore, this may explain the enhancement of DL SOT after oxygen exposure observed in this study. On the other hand, the larger enhancement of FL SOT (18.5 %) compared to DL SOT (8 %) indicates that the surface electric field has more evident effect on the FL SOT than DL SOT. The reduced increase rate of $\xi^{DL}$ and $\xi^{FL}$ at larger oxygen dose is attributed to the saturation of oxygen coverage on Co surface due to reduced sticking coefficient at larger dose. As mentioned above, it is difficult for the oxygen to diffuse into sub-surface layer of Co even with elongated oxygen exposure[43,50], which is also reflected by the smaller decrease of $M_s/M_s^0$ at larger oxygen dose shown in Figure 1c. To shed more lights on the oxygen exposure effects, we conducted the following control experiments: 1) dusting the oxygen exposed Co surface with a thin layer of Mg (~ 0.2 nm) to form Pt/Co/O/Mg and 2) capping the Pt/Co with a MgO layer, which is prepared by oxidizing 1 nm Mg layer with an oxygen dose of 2800 L. All the processes are performed in the UHV system without breaking the vacuum. We then performed the AHE and SOT measurements on both samples. The AHE loops of Pt/Co, Pt/Co/O (312 L), Pt/Co/O (312 L)/Mg and Pt/Co/MgO are shown in Figure 5a, from which it can be seen that all the samples present good PMA, though the nucleation field and $\Delta R_{AHE}$ of Pt/Co/O are smaller than the other three structures, which are caused by decreased magnetic moment in Co after oxygen exposure. As we reported previously, the Mg dusting is able to restore partially the quenched magnetic moment because in this case the charge transfer from Co to oxygen will be reduced as the latter can obtain some charges from Mg as well[30]. Figures 5b and 5c compare the DL and FL SOT efficiency for Pt/Co, Pt/Co/O, Pt/Co/O/Mg and Pt/Co/MgO structures (note: all of these structures are formed in sequence on the same sample instead of four different samples). Compared to Pt/Co, both DL and FL SOT in Pt/Co/MgO are enhanced by about 24 %, which is larger than the enhancement obtained by oxygen exposure, while the SOT in Pt/Co/O/Mg is even smaller than that in Pt/Co. This result demonstrates that the oxide capping layer is more efficient than oxygen in enhancing the SOT, and the interface of Co/oxide plays an important role in the SOT enhancement. This is in agreement with our previous findings that the Co-O-Mg bonding is the key contributor to PMA instead of Co-O bonding. Since the MgO can affect PMA, it implies that it can also change the local electric field at the Co/MgO interface and thus enhance the SOT. On the other hand, although the Co-O bonding can also enhance the SOT to a certain extent, it suppresses the saturation magnetization. The decrease of SOT after the deposition of Mg dusting layer on Pt/Co/O might result from the decreased surface electric field due to compensation of Co electrons from Mg atoms, which also explains the recovery of magnetic moment in Co by Mg dusting[30].

To have a more quantitative understanding of the role of oxygen in REE for Pt/Co, we perform *ab initio* calculations. Pristine Pt/Co is studied first as a reference. Figure 6c displays the calculated band structure of Pt (4 ML)/Co (4 ML) (ML represents monolayer) with SOC, whose optimized structure is shown in Figure 6a. The high-symmetric points of the Brillouin zone are labelled as -X and X, and the red and blue symbols indicate two opposite spin directions of the electrons in Figure 6c. As can be seen, there is no obvious Rashba splitting present in the band structure. Though the band states between -0.4 eV and -0.3 eV show similar dispersion as Rashba splitting along direction of $\Gamma \rightarrow$ X, at the other direction of $\Gamma \rightarrow$ -X the Rashba splitting is absent, which may indicate that the REE at Pt/Co interface and Co surface is weak. Afterwards, oxygen atoms are added on Co surface as depicted in Figure 6b. Based on the calculation results of structural relaxation, the oxygen atoms are adsorbed at the *hcp* hollow sites on the Co surface, which is consistent with the reported first-principles calculations[51] and experimental (scanning tunnelling microscope (STM), low-energy electron-diffraction (LEED)) studies[52]. As discussed above, during the oxygen exposure the oxygen atoms are mostly adsorbed on the surface layer of Co at room temperature and low oxygen partial pressure. As oxygen dose increases, the oxygen atoms are likely to form an oxide first with the surface layer of Co instead of diffusing into sub-surface layer due to the high energy barrier for diffusion[43,50]. Therefore, in the calculation the maximum oxygen coverage is set to be 1 ML, corresponding to the oxygen dose used in the experiments.

The calculated band structure of Pt (4 ML)/Co (4 ML)/O (1 ML), as shown in Figure 6d, displays evident Rashba splitting between -0.4 eV and -0.35 eV around Γ point. This is consistent with our experimental result that oxygen exposure on Co surface increases SOT efficiency resulting from the enhanced REE at Co/O surface. To estimate its Rashba parameter, we replot the band dispersion of Pt (4 ML)/Co (4 ML)/O (1 ML) in Figure 6e, from which the extracted $E_0$ and $k_0$ as indicated in the plot are 3.03 meV and 0.0136 Å$^{-1}$ respectively. By using the equation $\alpha_R = 2E_0/k_0$ and the extracted values of $E_0$ and $k_0$, we obtain $\alpha_R$ = 446 meV Å. This value is comparable with the reported $\alpha_R$ in structures with large REE[53-55], indicating a significant role of surface-adsorbed oxygen in enhancing the REE, thereby increasing the SOT in Pt/Co bilayers. Phenomenologically, the enhancement can be considered as originated from the large local electric field at Co/O surface due to the charge transfer from Co to O as shown schematically in Figure 6f. As we reported previously, an average of 0.84 *e* charge is transferred from Co to oxygen[30], which will result in a very large electric field considering the small distance between Co and oxygen.

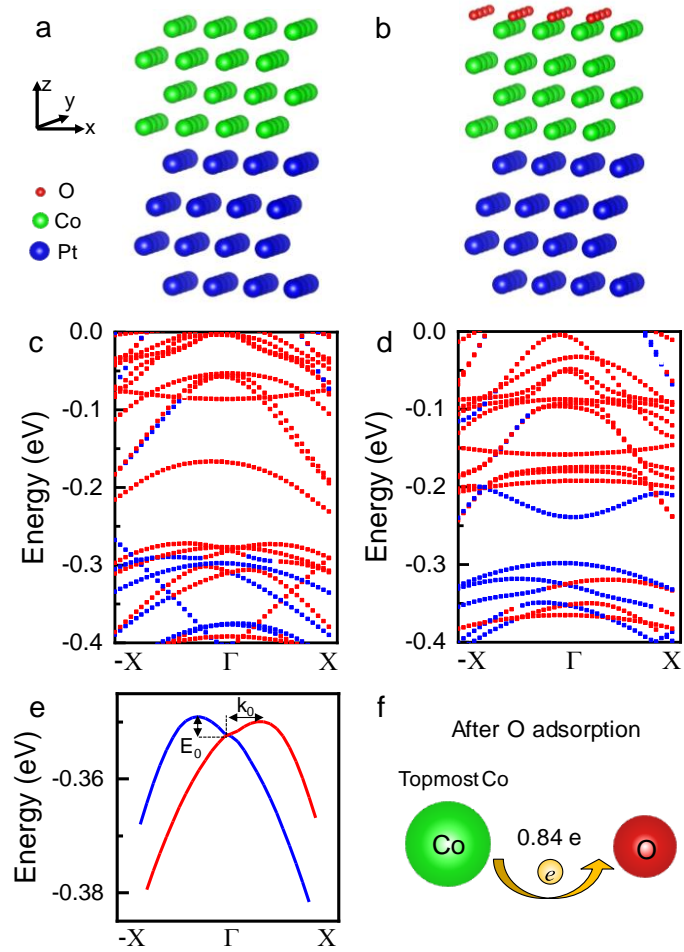

Figure 6. Schematic for optimized structures used in *ab initio* calculations of (a) Pt/Co and (b) Pt/Co/O, where Pt, Co and O are represented by blue, green and red balls, respectively. Calculated band structures for (c) Pt (4 ML)/Co (4 ML) and (d) Pt (4 ML)/Co (4 ML)/O (1 ML) with SOC, the red and blue symbols indicate two opposite spin directions of the electrons. (e) Zoom-in of the band dispersion in (d) for calculation of Rashba parameter. (f) Schematic of charge transfer from Co to O.

## Conclusion

In conclusion, we have carried out a systematic study of the oxygen exposure effect on spin-orbit torques of Pt/Co through *in-situ* deposition and measurement in an UHV environment. We found that the spin-orbit torque increases at low oxygen dose and gradually saturates at high dose. The size of enhancement is smaller than those reported in literature. The MgO capping layer has a similar effect with the magnitude of enhancement larger than that of surface-adsorbed oxygen. The results of *ab initio* calculations suggest that the enhancement of SOT by oxygen exposure is due to the enhanced Rashba-Edelstein effect, which is in turn caused by charge transfer from Co to oxygen. Our work sheds some light on the varying roles of oxygen in SOT engineering reported so far by different groups in a variety of material systems, and points to the importance of oxide capping layer in SOT-based devices.

## Data Availability

The datasets generated during and/or analysed during the current study are available from the corresponding author on reasonable request.

## Acknowledgements

Y.H.W. would like to acknowledge support by the Ministry of Education, Singapore under its Tier 2 Grant (Grant No. MOE2017-T2-2-011). Y.H.W. is a member of the Singapore Spintronics Consortium (SG-SPIN).

## Author Contributions

Y.H.W. designed and supervised the project. H.X. prepared the samples and conducted the experiments. J.R.Y. performed the *ab initio* calculations. Z.Y.L. and Y.M.Y. helped with the measurements. H.X., Z.Y.L. and Y.H.W. analysed and interpreted the results. H.X., J.R.Y. and Y.H.W. wrote the manuscript. All authors contributed to the final version of the manuscript.

## Additional Information

**Competing Interests:** The authors declare no competing financial interest.

**Figure Legends**

Figure 1. (a) Schematic of the UHV system for *in-situ* characterization and deposition, which consists of a nano-probe chamber, a transfer chamber and a deposition chamber. The inset shows the schematic of Hall bar used for SOT characterization and the sample structure. (b) AHE loop for Pt (2)/Co (0.8) measured with a perpendicular magnetic field $H_z$, where $+M_z$ and $-M_z$ indicate the up and down magnetization. (c) Normalized saturation magnetization $M_s/M_s^0$ versus oxygen dose.

Figure 2. First harmonic Hall voltages of Pt (2)/Co (0.8) with applied in-plane field of (a) $H_x$ and (b) $H_y$, and second harmonic Hall voltages with applied in-plane field of (c) $H_x$ and (d) $H_y$ measured at an AC current with an amplitude of 2.4 mA. $+M_z$ and $-M_z$ indicate the up and down magnetization.

Figure 3. (a) Damping-like SOT effective field $H^{DL}$ and (b) field-like SOT effective field $H^{FL}$ as a function of current density $J$ at oxygen dose of 0 L (circle), 30 L (triangle) and 312 L (square); the lines are linear fittings of $H^{DL}$ versus $J$. (c) $H^{DL}/J$ and (d) $H^{FL}/J$ as a function of oxygen dose.

Figure 4. Damping-like SOT efficiency $\xi^{DL}$ (square, left axis) and field-like SOT efficiency $\xi^{FL}$ (circle, right axis) versus oxygen dose.

Figure 5. (a) AHE loops, (b) $\xi^{DL}$ and (c) $\xi^{FL}$ of Pt/Co, Pt/Co/O (312 L), Pt/Co/O (312 L)/Mg and Pt/Co/MgO, where Pt layer is omitted in the legend in the figures.

Figure 6. Schematic for optimized structures used in *ab initio* calculations of (a) Pt/Co and (b) Pt/Co/O, where Pt, Co and O are represented by blue, green and red balls, respectively. Calculated band structures for (c) Pt (4 ML)/Co (4 ML) and (d) Pt (4 ML)/Co (4 ML)/O (1 ML) with SOC, the red and blue symbols indicate two opposite spin directions of the electrons. (e) Zoom-in of the band dispersion in (d) for calculation of Rashba parameter. (f) Schematic of charge transfer from Co to O.